\documentclass[prd,preprint,tightenlines,amsfonts,amsmath,amssymb]{revtex4}
\usepackage[pdftex]{graphicx}
\usepackage{float,color}
\usepackage{physics}

\begin{document}
\title{Qubit clock in quantum cosmology}
\author{Yasusada Nambu}
\email{nambu@gravity.phys.nagoya-u.ac.jp}

\affiliation{Department of Physics, Graduate School of Science, Nagoya
University, Chikusa, Nagoya 464-8602, Japan}

\begin{abstract}
  We investigate the emergent time scenario in quantum
  cosmology based on the Page-Wotters approach. Using a quantum
  cosmological model with a qubit clock, it is demonstrated how the
  entanglement between the qubit clock and the geometry derives
  emergence of a time parameter which defines evolution of the
  timeless quantum state of the universe. We show the universe wave
  function conditioned by a qubit clock obeys the standard
  Schr\"{o}dinger equation and the Fisher information for the clock
  state, which quantifies entanglement between the universe and the
  clock, contributes as a negative energy density.
\end{abstract}

\date{February 3, 2022}

\maketitle

  \newcommand{\anno}[1]{\textcolor{black}{#1}}

\section{Introduction} 

In the canonical approach to quantum mechanics, one supposes
existence of classical time as external parameter used by the
observer. The quantum state of the target system obeys the time
dependent Schr\"{o}dinger equation. The apparent lack of external time
for the dynamics of canonical quantum gravity, described by the
so-called ``problem of time'', is represented by the Wheeler-DeWitt (\anno{WDW})
equation~\cite{Dewitt1967}
\begin{equation}
   \hat H\ket{\Psi}=0,
\end{equation}
for the quantum state of the universe $\ket{\Psi}$. \anno{This
  equation implies the diffeomorphism invariance of the physical
  state, and owing to this structure, the physical state loses a
  concept of evolution with respect to a time parameter. Among
  several approaches to resolve the problem of time \cite{Kiefer2012},
  one idea of extracting dynamics from a seemingly stationary system
  has been proposed called the Page-Wotters approach or the
  conditional probability interpretation \cite{Melorose2015,
    Wootters1984}.}  \anno{Let us consider a bipartite system composed
  of  an universe
  system with the Hamiltonian $\hat H_U$ and a clock system with the Hamiltonian $\hat H_C$. }In the Page-Wotters
approach, ``time'' evolution is read, under the condition that the
total state Hamiltonian
\begin{equation}
    \hat H=\hat H_U\otimes\mathbb{I}_C+\mathbb{I}_U\otimes\hat H_C
\end{equation}
is constrained, in the quantum correlation between bipartitions of the
total system, the universe state (U) and the clock state (C) entangled
with it. Time evolution emerges from the measurement of an observable
clock of some kind, whose reading conditions the physical state. 

In our previous study \cite{Rotondo2019}, this scenario was
investigated using a Friedman-Lemaitre-Robertson-Walker (FLRW) model
with a minimal scalar field. \anno{Regarding the energy eigenstate of
  the scalar field as a system of quantum clock, the wave function of
  the universe conditioned by the clock is obtained.  The effect of
  the geometry (universe) on the clock state is encoded in the Berry
  phase of the scalar field state. We investigated the condition for
  the emergence of the standard Schr\"{o}dinger equation for the
  universe wave function conditioned by the scalar clock. The equation
  contains the Berry connection, which reflects strength of a coupling
  between the universe and the clock, and its behavior is also related
  to emergence of time and entanglement between the universe and the
  clock. We considered several cases of the clock states with
  superposition of energy eigenstates and discussed a condition of
  workable clock. However, we did not investigate so much about
  entanglement property in the Page-Wotters approach.}  In this
article, we revisit this problem using a qubit as a quantum clock in
cosmology. This model is simpler than the scalar field model and
captures an essential feature of the Page-Wooters mechanism. We can
explicitly evaluate entanglement between the qubit clock and the
universe, and discuss relation between the emergence of time and the
clock-system entanglement in the timeless quantum cosmology.

The reminder of this paper is organized as follows. In Section 2, we
introduce our quantum cosmological model with a qubit. In Section 3,
review how the qubit operates as a quantum clock. In Section 4, we
demonstrate how the qubit clock defines the time in the universe, and
relation to the clock-universe entanglement is discussed in Section
5. Section 6 in devoted to summary. We adopt units of $c=\hbar=1$
throughout this paper.

\section{Mini-superspace model with a qubit}
We consider a spatially flat FLRW universe with a qubit clock. The metric is
\begin{equation}
    ds^2=-N(t)^2dt^2+a(t)^2\delta_{ij}dx^idx^j,
\end{equation}
where $N(t)$ is the lapse function and $a(t)$ is the scalar factor. We
introduce the physical volume of the universe $\rho(t)=a^3(t)V$ as a
dynamical variable, where $V$ denotes the comoving
volume.  The total Hamiltonian is written as
\begin{equation}
    H=N(H_U\otimes\mathbb{I}_C+\mathbb{I}_U\otimes H_C),
\end{equation}
where
\begin{equation}
    H_U:=\rho\left(-\frac{3\kappa^2}{4}p_\rho^2+\frac{\Lambda}{\kappa^2}\right),
    \quad
    H_C:=
    \frac{\Omega}{2}\biggl(\ket{0}\bra{1}+\ket{1}\bra{0}\biggr),
\end{equation}
with $\kappa^2=8\pi G=1/M_p^2$.  $H_U$ is the Hamiltonian for the flat
FLRW universe with the spatial volume $\rho$ and the cosmological
constant $\Lambda$. $H_C$ is the Hamiltonian for the qubit with two
internal energy eigenstates $\{\ket{0},\ket{1}\}$ with the energy gap
$\Omega$. The total Hamiltonian is constrained to be zero
$\hat H_U\otimes\mathbb{I}_C+\mathbb{I}_U\otimes \hat H_C=0$, and the
physical state of the total system \anno{$\ket{\Psi}\rangle$} obeys the
\anno{WDW} equation:
\begin{equation}
\left[\hat\rho\,(-\hat p_\rho^2+\lambda)+\frac{\omega}{2}
    \bigl(\ket{0}\bra{1}+\ket{1}\bra{0}\bigr)\right]\anno{\ket{\Psi}\rangle}=0,
\end{equation}
where $\lambda=4\Lambda/(3\kappa^4)$ and
$\omega=2\Omega/(3\kappa^2)$. \anno{Here, to make clear the structure
  of the total state, we adopted the notation $\ket{\Psi}\rangle$ for
  a pure state in a bipartite Hilbert space, which is the tensor
  product of the universe Hilbert space and the qubit Hilbert space: }
\begin{equation}
  \anno{\ket{\Psi}\rangle:}=\ket{\Psi_{0}}\otimes\ket{0}+\ket{\Psi_{1}}\otimes\ket{1}.
  \label{eq:total-state}
\end{equation}
\anno{Let us introduce the wave function of the universe
  $\Psi_{0,1}(\rho):=\bra{\rho}\ket{\Psi_{0,1}}$ with the eigen-vector
  $\ket{\rho}$ of $\hat\rho$. Then the wave function obeys}
\begin{align}
  &\rho(\partial_\rho^2+\lambda)\Psi_0+\frac{\omega}{2}\Psi_1=0,\\
  &\rho(\partial_\rho^2+\lambda)\Psi_1+\frac{\omega}{2}\Psi_0=0.
\end{align}
For $\Psi_{\pm}:=\Psi_0\pm\Psi_1$,
\begin{equation}
    \left[\rho(\partial_\rho^2+\lambda)\pm\frac{\omega}{2}\right]\Psi_{\pm}=0.
    \label{eq:wdpm}
\end{equation}
The solution which is regular at $\rho=0$ is
\begin{equation}
  \Psi_\pm(\rho)=e^{-i\lambda^{1/2}\rho}\rho
  F[1+i\epsilon_\pm/(2\lambda^{1/2}),2,2i\lambda^{1/2}\rho],\quad
  \epsilon_{\pm}=\pm\frac{\omega}{2}, 
\end{equation}
where $F[\alpha,\gamma,z]$ is the confluence hypergeometric function and
its asymptotic behavior for $z\rightarrow\infty$ is
\begin{equation}
  F[\alpha,\gamma,z]\sim\frac{\Gamma(\gamma)}{\Gamma(\gamma-\alpha)}(-z)^{-\alpha}+\frac{\Gamma(\gamma)}{\Gamma(\alpha)}e^zz^{\alpha-\gamma}.
\end{equation}
Applying this formula, asymptotic form of the wave function
$\Psi_\pm(\rho)$ for $\rho\rightarrow\infty$ is given by
\begin{equation}
  \Psi_\pm\sim\frac{-1}{2i\lambda^{1/2}}\left\{\frac{(-2i\lambda^{1/2}
      \rho)^{-i\epsilon_\pm/(2\lambda^{1/2})}e^{-i\lambda^{1/2}\rho}}{\Gamma(1-i\epsilon_\pm/
      (2\lambda^{1/2}))}-\frac{(+2i\lambda^{1/2}\rho)^{+i\epsilon_\pm/(2\lambda^{1/2})}e^{+i\lambda^{1/2}\rho}}{\Gamma(1+i\epsilon_\pm/(2\lambda^{1/2}))}\right\}.
\end{equation}
From the behavior of phase factors of the solution,
the wave function represents superposition of expanding and
contracting universes with the qubit energy $\pm\omega/2$. From the \anno{WDW}
Eq.~\eqref{eq:wdpm}, the contribution of the qubit to the total matter
energy density is proportional to $\rho^{-1}$, which is the same
behavior as a dust fluid.

\section{Clock dynamics}
Let us first consider the clock dynamics of which Hamiltonian is given
by $H_C$.  Using the energy eigen basis $\{\ket{0},\ket{1}\}$ of the
qubit, the clock state is represented as
\begin{equation}
  \ket{\chi_C}=c_0(T)\ket{0}+c_1(T)\ket{1},\quad |c_0|^2+|c_1|^2=1,
  \label{eq:cstate1}
\end{equation}
where we assume that the evolution of the clock state is parameterized
by a time parameter $T$, and the state \anno{is assumed to} obey the
standard Schr\"{o}dinger equation:
\begin{equation}
    i\frac{\partial}{\partial T}\ket{\chi_C}=\hat H_C\ket{\chi_C}.
    \label{eq:shroe}
\end{equation}
For the state \eqref{eq:cstate1},
\begin{equation}
  i\,\dot c_0=\frac{\omega}{2}\,c_1,\quad i\,\dot
  c_1=\frac{\omega}{2}\,c_0,\quad \cdot{}=\frac{\partial}{\partial T}\quad,
  \label{eq:ceq}
\end{equation}
and the solution is
\begin{equation}
  c_0=\frac{1}{\sqrt{2}}\left(\alpha\, e^{-i\omega T/2}+\beta\, e^{i\omega
       T/2}\right),\quad
  c_1=\frac{1}{\sqrt{2}}\left(\alpha\, e^{-i\omega T/2}-\beta\, e^{i\omega T/2}\right),
\end{equation}
where $\alpha$ and $\beta$ are constants with $|\alpha|^2+|\beta|^2=1$.
Probabilities to realize states $\ket{0}$ and $\ket{1}$ are
\begin{align}
    P_{0}&=|c_{0}|^2
    =\frac{1}{2}\left[1+(\alpha\beta^* e^{-i\omega
             T}+\text{(c.c.)})\right],\\
  P_{1}&=|c_{1}|^2
    =\frac{1}{2}\left[1-(\alpha\beta^* e^{-i\omega
             T}+\text{(c.c.)})\right].
\end{align}
These probabilities oscillate with a period $2\pi/\omega$ and the
amplitude $|\alpha\beta|=|\alpha|\sqrt{1-|\alpha|^2}$. This oscillatory
behavior reflects superposition of two energy eigenstates $\ket{0}$
and $\ket{1}$ (interference visibility), and plays a role of a
``clock''. As a special case of the state, a state with $\alpha=0$ or
$\alpha=1$ corresponds to the energy eigenstate; the visibility becomes
zero and the system does not work as a clock because we cannot measure
the oscillatory behavior of the clock state.

In the cosmological situation \anno{where the clock is coupled to the
  external universe, constants $\alpha,\beta$ can be functions
  of the spatial volume of the universe $\rho$. We assume the
  following $\rho$ dependence in $c_0, c_1$:
  \begin{align}
     &c_0=\frac{1}{\sqrt{2}}\left(|\alpha|\, e^{-i\theta(\rho)/2}\,
       e^{-i\omega T/2}+|\beta|\, e^{i\theta(\rho)/2}\, e^{i\omega
       T/2}\right), \label{eq:csol1}\\
  &c_1=\frac{1}{\sqrt{2}}\left(|\alpha|\, e^{-i\theta(\rho)/2}\,
    e^{-i\omega T/2}-|\beta|\, e^{i\theta(\rho)/2}\, e^{i\omega
    T/2}\right), \label{eq:csol2} 
  \end{align}
  where $\theta(\rho)$ is an arbitrary function representing a
  relative phase between $\alpha$ and $\beta$.  The clock state is
  parameterized by $\rho$ as
\begin{equation}
    \ket{\chi_C(\rho)}=c_0(T,\theta(\rho))\ket{0}+c_1(T,\theta(\rho))\ket{1}.
    \label{eq:cstate}
\end{equation}
}
Regarding $\rho$ as an external parameter, the quantum Fisher
information \cite{PARIS2009,Li2013} of this state is given by
\begin{align}
  \mathcal{F}_Q(\rho):
  &
    =4\left[\bra{\partial_\rho\chi_C}\ket{\partial_\rho\chi_C}-\bra{\chi_C}\ket{\partial_\rho\chi_C}\bra{\partial_\rho\chi_C}\ket{\chi_C}\right]
    \label{eq:Fisher}\\
  &=4\biggl[\sum_{j=0,1}c_j'{}^*c_j'-\biggl(\sum_{j=0,1}c_j^*c_j'\biggr)\biggl(\sum_{j=0,1}c_j'^*c_j\biggr)\biggr]  \notag \\
  &=\anno{(\theta')^2}(c_0^2-c_1^2)(c_0^*{}^2-c_1^*{}^2) \notag \\
  &=\anno{(\theta')^2}|\alpha|^2|\beta|^2,
    \label{eq:Fisher2}
\end{align}
where $'=\partial/\partial\rho$. This quantity is related to the
distance or overlap between two states $\ket{\chi_C(\rho)}$ and
$\ket{\chi_C(\rho+d\rho)}$. The Bures distance between clock states is
defined by
\begin{align}
  D_Q(\rho):&=1-|\bra{\chi_C(\rho)}\ket{\chi_C(\rho+d\rho)}|^2 \notag
  \\
  &=
  \frac{\mathcal{F}_Q}{4}(d\rho)^2 +O(d\rho^3).
  \label{eq:distance}
\end{align}
The distance between two states is proportional to the Fisher
information, and becomes zero for $\mathcal{F}_Q=0$.  Figure 1 shows
the behavior of the quantum Fisher information as a function of
$|\alpha|^2$ for $\anno{\theta'}\neq0$. $|\alpha|=0,1$ corresponds to the
energy eigenstate and the Fisher information is zero. \anno{Figure 1 shows
$|\alpha|^2$ dependence of the Fisher information.}
\begin{figure}[H]
    \centering
    \includegraphics[width=0.4\linewidth,clip]{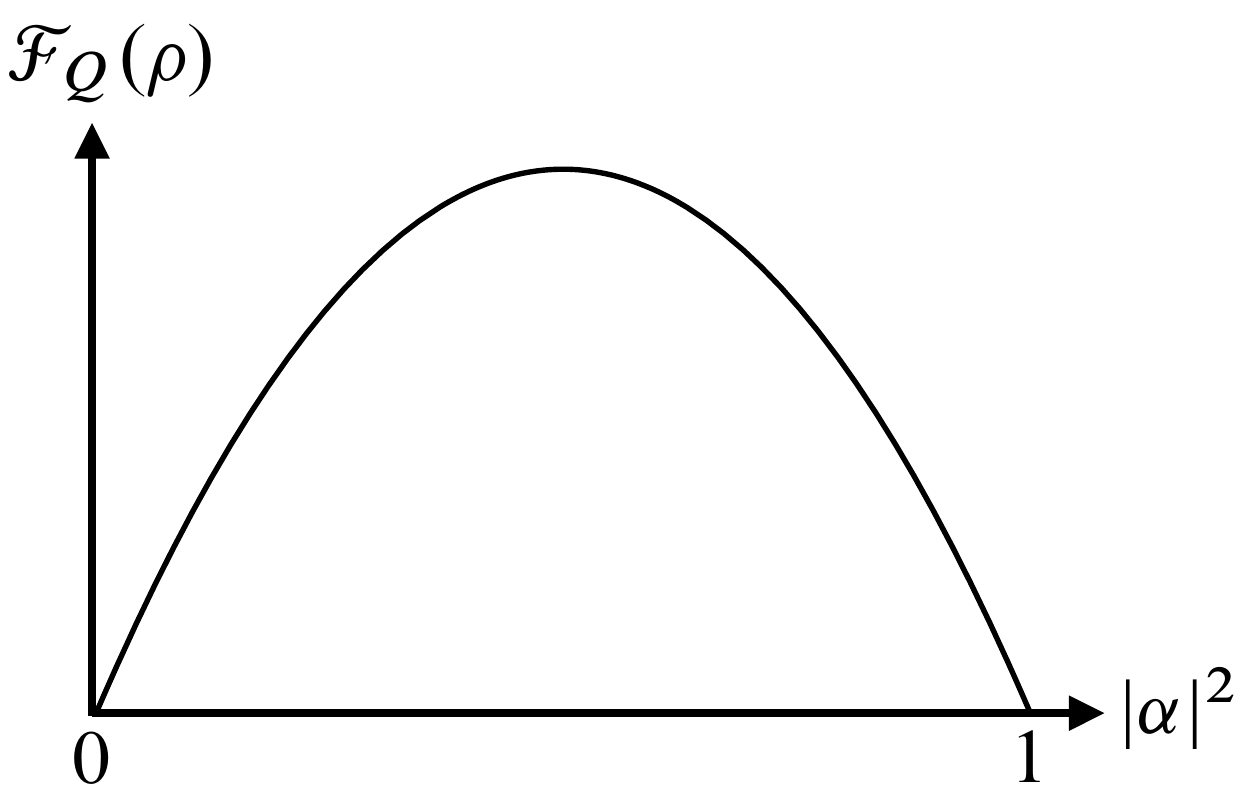}
    \caption{Behavior of the qunatum Fisher information as a function of a 
      $|\alpha|^2$. For $|\alpha|=0,1$, the state is the energy eigenstate and  the
      Fisher information becomes zero.}
\end{figure}
The Fisher information provides the Gramer-Rao bound
\cite{Braunstein1994} which gives a lower bound for the variance of
the estimation of the external parameter $\rho$:
\begin{equation}
  \expval{(\Delta\rho)^2}\ge \frac{1}{\mathcal{F}_Q(\rho)}.
\end{equation}
On the other hand, using the Schr\"{o}dinger equation
\eqref{eq:shroe}, the variance of the clock energy is expressed as
\begin{equation}
  \expval{(\Delta
    E)^2}:=\bra{\chi_C}\hat H_C^2\ket{\chi_C}-\bra{\chi_C}\hat H_C\ket{\chi_C}^2=\anno{\frac{\mathcal{F}_Q\,\omega^2}{4(\theta')^2}},
\end{equation}
and the Cramer-Rao inequality becomes
$\expval{(\Delta E)^2}\expval{(\Delta\rho)^2}\anno{
  (\theta')^2/\omega^2}\ge 1/4$. Thus, by introducing a quantity
$\expval{(\Delta
  T)^2}:=\expval{(\Delta\rho)^2}\anno{(\theta')^2/\omega^2}$, we
obtain the following uncertainty relation between $\Delta E$ and
$\Delta T$:
\begin{equation}
  \expval{(\Delta E)^2}\expval{(\Delta T)^2}\ge \frac{1}{4}.
\end{equation}
The variance $\expval{(\Delta T)^2}$ represents the accuracy of the
clock reading. If the clock state is an energy eigenstate, then
$\expval{(\Delta E)^2}=0$ and $\expval{(\Delta T)^2}$ becomes
infinite, which means that the qubit system does not properly operate
as a clock.  Thus for $\anno{\theta'}\neq 0$, the requirement for workable clock
is $\mathcal{F}_Q\neq 0$, or equivalently
\begin{equation}
    |\alpha\beta|\neq 0. \label{eq:wclock}
\end{equation}
This is the same condition that the state is not energy
eigenstate~\cite{Anandan1990}. The maximum accuracy of the clock is
achieved for $|\alpha|^2=1/2$.

\section{Quantum cosmology with a qubit clock}
To investigate how the qubit clock introduces a time parameter in the
universe, we define a wave function conditioned by the
clock state \eqref{eq:cstate}:
\begin{equation}
  \psi(\rho,T):=\anno{(\bra{\rho}\otimes\bra{\chi_C})\ket{\Psi}\rangle}=c_0^*\,\Psi_{0}+c_1^*\,\Psi_{1},
  \label{eq:cstate}
\end{equation}
where $c_0=c_0(T,\anno{\theta(\rho)})$ and
$c_1=c_1(T,\anno{\theta(\rho)})$.
Then using the \anno{WDW} equation \eqref{eq:wdpm}, it can
be shown that the wave function $\psi$ obeys the following equation:
\begin{align}
  &\psi''+\lambda\psi+\frac{\omega}{2\rho}(c_0^*\Psi_1+c_1^*\Psi_0)=
    -\sum_{j=0,1}c_j^*(2c_j{}'\psi'
    +c_j{}''\psi).
\end{align}
\anno{It is possible to rewrite this equation to the following form:
  \begin{equation}
    \left[\hat
      D^2+\lambda\right]\psi+\frac{\omega}{2\rho}(c_0^*\Psi_1+c_1^*\Psi_0)=
   -\bra{\chi_C}\ket{\hat D^*{}^2\chi_C}\psi, \label{eq:WDg1}
  \end{equation}
  where the covariant derivative is introduced as
  $\hat D:=\partial_\rho-iA, \hat D^*:=\partial_\rho+iA^*$ with the Berry connection
  \begin{equation}
    A:=i\bra{\chi_C}\ket{\partial_\rho\chi_C}=i\sum_{j=0,1}c_j^*c_j'.
  \end{equation}
  In this case, the Berry connection represents impact of the external
  parameter $\rho$ on the clock state $\ket{\chi_C}$.  Equation
   \eqref{eq:WDg1} is invariant under the local gauge
  transformation
  $\psi\rightarrow \widetilde\psi=e^{if(\rho)}\psi,
  \ket{\chi_C}\rightarrow\ket{\widetilde\chi_C}=e^{-if(\rho)}\ket{\chi_C}$,
  which reflects the gauge invariance of the total state. Now we
  choose the ``zero connection'' gauge
  $\widetilde
  A=\bra{\widetilde\chi_C}\ket{\partial_\rho\widetilde\chi_C}=0$ to
  make the equation simpler. The condition for $f(\rho)$ of this gauge
  is
\begin{equation}
  if'=\bra{\chi_C}\ket{\partial_\rho\chi_C}=\sum_{j=0,1}c_j^*c'_j.
\end{equation}
The wave function $\widetilde\psi=e^{if(\rho)}\psi$ with the zero
connection gauge obyes
\begin{equation}
  \widetilde\psi''+\lambda\widetilde\psi
  +\frac{\omega}{2\rho}(c_0^*\,\Psi_1+c_1^*\,\Psi_0)=
  -\bra{\widetilde\chi_C}\ket{\partial_\rho^2\widetilde\chi_C}\widetilde\psi,
  \label{eq:WDg2}
\end{equation}
and the coefficent of the right-hand side of this equation is
\begin{equation}
  -\bra{\widetilde\chi_C}\ket{\partial_\rho^2\widetilde\chi_C}=
  \bra{\partial_\rho\chi_C}\ket{\partial_\rho\chi_C}-\bra{\chi_C}\ket{\partial_\rho\chi_C}\bra{\partial\chi_C}\ket{\chi_C}=\mathcal{F}_Q/4.
\end{equation}
}
\anno{After all,} the wave function $\widetilde\psi$ satisfies
\begin{equation}
  \widetilde\psi''+\lambda\widetilde\psi+\frac{\omega}{2\rho}e^{-if}\left(c_0^*\Psi_1+c_1^*\Psi_0\right)
=\frac{\mathcal{F}_Q(\rho)}{4}\widetilde\psi,
\label{eq:tpsi}
\end{equation}
where the quantum Fisher information of the clock state appears in the
right-hand side after the gauge transformation.
In the left-hand side of Eq.~\eqref{eq:tpsi}, using Eq.~\eqref{eq:ceq},
\begin{align}
    c_0^*\,e^{-if}\Psi_1+c_1^*\,e^{-if}\Psi_0
 & =-\frac{2i}{\omega}(
    \dot c_1^*\,e^{-if}\Psi_1+\dot
   c_0^*\,e^{-if}\Psi_0) \notag \\
  &=-\frac{2i}{\omega}\left(\frac{\partial\widetilde\psi}{\partial
    T}-\sum_{j=0,1}c_j^*\dot c_j\,\widetilde\psi\right),
\end{align}
\anno{where $\sum_jc_j^*\dot c_j=(\omega/2i)(|\alpha|^2-|\beta|^2)$
  using \eqref{eq:csol1} and \eqref{eq:csol2}.}
Therefore,  we obtain the following  Schr\"{o}dinger equation for the wave
function of the universe $\widetilde\psi$, which represents time
evolution of the wave function of the universe conditioned by the
qubit clock:
\begin{equation}
  i\frac{\partial\widetilde\psi}{\partial T}=\rho\left[\frac{\partial^2}{\partial\rho^2}
    +\left(\lambda+\frac{(|\alpha|^2-|\beta|^2)\omega}{2\rho}-\frac{\mathcal{F}_Q(\rho)}{4}\right)\right]\widetilde\psi.
  \label{eq:tsch}
\end{equation}
In deriving this equation, we should keep in mind that the condition
of workable clock \eqref{eq:wclock} is assumed; if this condition is
not fulfiled, the clock state becomes the energy eigenstate. Then the
$T$-dependence included in the phase of $\widetilde\psi$ disappears,
and relevant information of time evolution of the universe is lost. In
this case, the original timeless \anno{WDW} equation \eqref{eq:wdpm} is
recovered:
\begin{equation}
  \rho\left[\frac{\partial^2}{\partial\rho^2}+\left(\lambda\pm
    \frac{\omega}{2\rho}\right)\right]\widetilde\psi=0,
\end{equation}
where the plus and minus signs correspond to the positive
($|\alpha|=1, |\beta|=0$) and negative ($|\alpha|=0, |\beta|=1$)
energy eigenstate of the clock, respectively.

In the
right-hand side of Eq.~\eqref{eq:tsch}, $\omega/\rho$ represents the
contribution of the qubit to matter energy density, which shows the
same behavior as the classical dust fluid. On the other hand, the
Fisher information contributes as a matter field with negative energy
density because the Fisher information is positive semi-definite.

\section{Entanglement}
We evaluate entanglement between the qubit and the universe to confirm
 how the emergence of time due to the Page-Wotters mechanism is
related to entanglement.  As defined in \eqref{eq:cstate}, the total
state conditioned by the clock is written as
\begin{equation}
  \anno{\ket{\psi(\rho)}}=\psi(\rho)\ket{\chi_C(\rho)},\quad
  \ket{\chi_C(\rho)}=\sum_{j=0,1}c_j(\rho)\ket{j},\quad |c_0|^2+|c_1|^2=1.
\end{equation}
The density matrix of this state is
\begin{equation}
  \Xi(\rho_1,\rho_2):=\anno{\ket{\psi(\rho_1)}\bra{\psi(\rho_2)}}=\psi(\rho_1)\psi^*(\rho_2)\ket{\chi_C(\rho_1)}\bra{\chi_C(\rho_2)}.
\end{equation}
By tracing out the qubit degrees of freedom, the reduced state is
\begin{align}
  \Xi_\text{red}(\rho_1,\rho_2)&=\mathrm{Tr}_\text{qubit}\,\Xi(\rho_1,\rho_2)
  \notag \\
 &\anno{=\psi(\rho_1)\psi^*(\rho_2)\sum_{j=0,1}\bra{j}\ket{\chi_C(\rho_1)}\bra{\chi_C(\rho_2)}\ket{j}  }\notag \\
  &=\psi(\rho_1)\psi^*(\rho_2)\bra{\chi_C(\rho_2)}\ket{\chi_C(\rho_1)},
   \end{align}
   and
   \begin{equation}
       \mathrm{Tr}\,\Xi_\text{red}=\int d\rho|\psi(\rho)|^2,
   \end{equation}
   where we do not specify explicit normalization of the wave
   function. Then the purity is 
\begin{align}
  \mathrm{Tr}\,\Xi_\text{red}^2
  & =\int d\rho_1
    d\rho_2|\psi(\rho_1)|^2|\psi(\rho_2)|^2|\bra{\chi_C(\rho_1)}
    \ket{\chi_C(\rho_2)}|^2\le (\mathrm{Tr}\,\Xi_\text{red})^2, \label{eq:purity}
\end{align}
where the overlap of two clock states is written as
\anno{
\begin{equation}
    |\bra{\chi_C(\rho_1)}\ket{\chi_C(\rho_2)}|^2=\left|\sum_{j=0,1}c_j^*(\rho_1)
                                                  c_j(\rho_2)\right|^2
                                                  =1-4|\alpha|^2|\beta|^2\sin^2\left(\frac{\theta_{12}}{2}\right),
             \label{eq:overlap}                                     
\end{equation}
with $\theta_{12}=\theta(\rho_1)-\theta(\rho_2)$}.  Equality in
Eq.~\eqref{eq:purity} is attained when the reduced system is pure
state, in which case the qubit and the universe is separable (product
state).  As we have shown in \eqref{eq:Fisher2} and
\eqref{eq:distance}, the condition of
$\bra{\chi_C(\rho_1)} \ket{\chi_C(\rho_2)}=1$ for $\rho_1\neq \rho_2$
is equivalent to $\mathcal{F}_Q(\rho)=0$ and $\ket{\chi_C}$ is the
energy eigenstate. In this case,
$\mathrm{Tr}\,\Xi_\text{red}^2=(\mathrm{Tr}\,\Xi_\text{red})^2$ and
the reduced system becomes pure. This implies that there is no
entanglement between the qubit and the universe for
$\mathcal{F}_Q=0$. The qubit system does not work as a clock for the
universe.

On the other hand, for
\anno{$|\bra{\chi_C(\rho_1)}\ket{\chi_C(\rho_2)}|<1$} case,
$\mathrm{Tr}\,\Xi_\text{red}^2<(\mathrm{Tr}\,\Xi_\text{red})^2$ and
the reduced system becomes a mixed state. In this case, the qubit and
the universe are entangled, and the universe wave function is
parameterized by the time introduced by the qubit. \anno{The
  entanglement can be quantified using the mixedness, which is
  proportional to $|\alpha|^2|\beta|^2$ as shown by \eqref{eq:overlap}
  in the present model. As presented in Eq.~\eqref{eq:Fisher2}, the Fisher
  information shows the same $|\alpha|^2|\beta|^2$ dependence, the
  Fisher information is an quantifier of the entanglement between the
  clock and the universe in the present model.}  The maximal
entanglement is achieved for $|\alpha|^2=1/2$.

\section{Summary}
Using a qubit clock, we demonstrate the Page-Wootters mechanism in
quantum cosmology which explains emergence of time in timeless
stationary quantum systems. The essence of this mechanism is
entanglement between a clock degrees of freedom and the target system;
the information of the clock system is mediated by entanglement and
transfered to the timeless universe. We considered the condition of a
workable qubit clock and show that the condition is non-zero values of
the quantum Fisher information for the qubit state, which regards
volume of the universe as an external parameter.  As the
qubit-universe system is totally constrained, non-zero Fisher
information of the qubit system implies that the universe is capable
of acquiring information of the qubit system via entanglement between
the qubit and the universe. As shown in Eq.~\eqref{eq:tsch}, the
Fisher information  quantifies the strength of entanglement, and 
modefies the expansion law of the quantum universe because it serves
as an negative energy density of which $\rho$-dependence is determined
through $\anno{\theta(\rho)}$.

Recently, quantum nature of gravity is paid much attention from the
viewpoint of quantum information \cite{Bose2017,Marletto2017a}; by
regarding graviational interaction as a quantum channel, it inevitably
introduces entanglement between a probe system and a target system.
In the treatment of canonical quantization of gravity, owing to the
\anno{WDW} equation, entanglement between matter fields and gravity is
encoded as universal way, and this structure is independent of detail
of matter fields. \anno{Therefore,} the emergent time scenario in the canonical
quantum gravity is deeply related to quantum informational nature of
gravity.

\begin{acknowledgments}
  The author would like to thank Y. Kaku, S. Maeda and
  Y. Osawa for useful discussion on quantumness of gravity.
  Preliminary idea of this study was obtained through communication with them.
\end{acknowledgments}


\begin{thebibliography}{10}
\newcommand{\enquote}[1]{``#1''}

\bibitem{Dewitt1967}
DeWitt,~B.~S. \enquote{{Quantum theory of gravity. I. the canonical theory}},
  \emph{Phys. Rev.} \textbf{1967}, 160, 1113--1148.

\bibitem{Kiefer2012}
  Kiefer,~C. \enquote{Quantum Gravity},
  Oxfort University Press \textbf{2012}.
  
\bibitem{Melorose2015}
Page,~D.N.~;~Wootters,~W.~K. \enquote{{Evolution without evolution: Dynamics
  described by stationary observables}}, \emph{Phys. Rev. D}
\textbf{1983}, 27, 2885--2892.

\bibitem{Wootters1984}
Wootters,~W.~K. \enquote{{"Time" replaced by quantum correlations}}, \emph{Int.
  J. Theor. Phys.} \textbf{1984},  23, 701--711.

\bibitem{Rotondo2019}
Rotondo,~M.~;~Nambu,~Y. \enquote{{Clock Time in Quantum Cosmology}},
  \emph{Universe} \textbf{2019}, 5, 66.

\bibitem{PARIS2009}
Paris,~M.~G.~A. \enquote{{QUANTUM ESTIMATION FOR QUANTUM TECHNOLOGY}},
  \emph{Int. J. Quantum Inf.} \textbf{2009}, 07~(supp01), 125--137.

\bibitem{Li2013}
Li,~N.~;~Luo,~S. \enquote{{Entanglement detection via quantum Fisher
  information}}, \emph{Phys. Rev. A} \textbf{2013}, 88,  014301.

\bibitem{Braunstein1994}
Braunstein,~S.~L.~;~Caves,~C.~M. \enquote{{Statistical Distance and the
  Geometry of Quantum States}}, \emph{Phys. Rev. Lett.} \textbf{1994},
72,  3439--3443.

\bibitem{Anandan1990}
Anandan,~J.~; Aharonov,~Y. \enquote{{Geometry of quantum evolution}},
  \emph{Phys. Rev. Lett.} \textbf{1990}, 65, 1697--1700.

\bibitem{Bose2017}
Bose,~S.~;~Mazumdar,~A.~;~Morley,~G.~W.~;~Ulbricht,~H.~;~Toro{\v{s}},~M.~;
~Paternostro,~M.~;~Geraci,~A.~A.~;~Barker,~P.~F.~;Kim,~M.~S.~;~Milburn,~ G.
  \enquote{{Spin Entanglement Witness for Quantum Gravity}}, \emph{Phys. Rev.
  Lett.} \textbf{2017}, 119, 240401.

\bibitem{Marletto2017a}
Marletto,~C.~;~Vedral,~V. \enquote{{Witness gravity's quantum side in the
  lab}}, \emph{Nature} \textbf{2017}, 547, 156--158.

\end{thebibliography}
\end{document}